\documentclass[number,review,12pt]{elsarticle}
\usepackage{multirow}
\usepackage{epsfig, setspace, pmgraph, epstopdf}
\usepackage[lofdepth,lotdepth]{subfig}
\usepackage{graphicx}
\usepackage{afterpage}
\usepackage{amsmath, amsfonts, amssymb}
\usepackage{dsfont,courier}
\usepackage{verbatim}
\usepackage{upgreek}
\usepackage{longtable}
\usepackage{hyperref}
\usepackage{xcolor,soul}
\usepackage[modulo]{lineno}
\usepackage{caption}
\usepackage{booktabs}
\usepackage[margin=1 in]{geometry}%
\usepackage[numbers]{natbib}
\usepackage{pdfpages}
\usepackage[shortlabels]{enumitem}
\usepackage{calc}
\usepackage{blindtext}
\usepackage{pdfpages}
\modulolinenumbers[2]

\journal{Materials Science \& Engineering: A}

\begin{document}

\begin{frontmatter}

\title{Response of A356 to warm rotary forming and subsequent T6 heat treatment}

\author[uom]{M. J. Roy\corref{cor1}}
\ead{matthew.roy@manchester.ac.uk}
\author[ubc]{D. M. Maijer}
\ead{daan.maijer@ubc.ca}

\address[uom]{School of Mechanical, Aerospace and Civil Engineering, The University of Manchester, Manchester, United Kingdom M13 9PL}
\address[ubc]{Dept. of Materials Engineering, The University of British Columbia, Vancouver, BC, Canada V6T 1Z4}

\cortext[cor1]{Corresponding author, Tel. +44 161 275 1916}

\begin{abstract}
The through-process microstructural effects in A356 subjected to rotary forming at elevated temperatures have been investigated. Macro and micro-hardness testing have been used extensively to track changes in the material from the as-cast state to as-formed, and T6 heat treated. Targeted thermal treatments have been used to isolate the effects of mechanical deformation through comparative measurements. These measurements include macro and micro hardness measurements, Energy-dispersive X-ray analysis and examination of eutectic-Si particle size and morphology. The results indicate that the as-cast material is stable up to approximately 144$^\circ$C, with the rotary formed material exhibiting decreased macrohardness in-line with the time spent at elevated temperature. Post heat treatment, there was a significant decrease in hardness with increased levels of deformation. Results indicate that precipitation hardening is not appreciably affected by rotary forming, and the principal cause for the drop in hardness with deformation is due to the condition of Al-Si eutectic phase.
\end{abstract}

\begin{keyword}
A356 \sep microstructure \sep eutectic \sep flow forming \sep deformation \sep heat treatment
\end{keyword}

\end{frontmatter}
\begin{linenumbers}
\section{Introduction}\label{sec:intro}
The use of Al-Si-Mg alloy castings enables the manufacture of near-net shape, lightweight components for many industries. However, design considerations must be made to account for casting inhomogeneity and porosity that may limit service life \cite{Roy.12b}. Forging the casting during solidification \cite{Ashouri.08} or through rheocasting \cite{Lashkari.08} is one method for improving mechanical properties. These processes typically involve prohibitively high forming loads, accompanied by high operational costs. An alternative to minimize these negative aspects for axisymmetric components is through rotary forming.

Rotary forming is a general term to describe similar forming techniques such as spinning, shear/flow forming, which are  incremental forming techniques employed on circular or tubular workpieces. The workpieces, or blanks, are attached to a mandrel and are spun into contact with an impinging roller or stationary tool which locally plasticizes the material and induces it to move axially and radially. A general review of spinning processes has been conducted by Wong et al. \cite{Wong.03}. Music et al. \cite{Music.10} recently conducted an extensive review specifically devoted to process mechanics. Experimental studies of this process conducted at ambient temperatures on wrought aluminum alloys have demonstrated that large amounts of plastic deformation may be imparted. Haghshenas et al. \cite{Haghshenas.12} reported that equivalent strains of up to 1.2 may be imparted to Al 6061-O workpieces, and 1.7 for steel workpieces \cite{Haghshenas.11}.  Applications of this forming technique specifically to cast aluminum alloys have shown the potential to reduce or eliminate porosity and thereby significantly improving fatigue performance. However, due to the lack of ambient ductility, spinning of cast aluminum alloys requires deformation at elevated temperatures in order to achieve a sound product.

Mori et al. \cite{Mori.09} conducted spinning experiments on cast A357 alloy blanks machined from larger castings at temperatures between 350 and 400$^{\circ}$C. Post-deformation analysis found that porosity had been eliminated for wall thickness reductions of 25\% and greater. While not quantified, it was reported that the dendrite arm spacing (DAS) was reduced in-line with the wall reduction level. As compared to unformed material, it was reported than the yield strength of the formed material increased after T6 heat treatment (solution treatment at 545$^{\circ}$C for 4 hours, followed by ageing for 8 hours at 175$^{\circ}$C). Ductility, as characterized by elongation, was increased by approximately twofold over all deformation levels. Furthermore, the as-deformed material was not characterized prior to heat treatment and it is therefore difficult to differentiate between the effects on strength due to deformation and heat treatment.

Zhao et al. \cite{Zhao.11} conducted elevated temperature spinning experiments on strontium modified, low pressure die cast A356 tubes with a starting wall thickness of 23.0 mm. At severe wall thickness reductions, the dendritic structure was no longer recognizable in some locations. Average dendrite arm spacing was modified from 37.2 to 23.0 $\upmu$m at wall thickness reductions of 80\%. Mechanical testing of this material also showed improvements in the mechanical properties following heat treatment. It was reported that the Brinell hardness increased by $\sim$14\% with a 70\% wall thickness reduction. The hardness reported in the unformed condition matched that of Tash et al. \cite{Tash.07} for material in the solutionized condition suggesting that a non-standard heat treatment was employed. Zhao et al. did not disclose the forming temperature employed.

Cheng et al. \cite{Cheng.10} employed a numerically controlled industrial forming apparatus to reduce wall thicknesses of A356 blanks with a diameter of $\sim$400 mm and a starting wall thickness of $\sim$8 mm. Maximum thickness reduction was reported as 60\%. Cheng et al. found the same effects on microstructure as the two previous studies with a processing temperature of 350$^{\circ}$C, however, they also reported a small decrease in Rockwell hardness of the material in the spun condition ($H_{\mathrm{RF}}=90.5 \pm 1.5$ versus $89.3 \pm 0.7$) post solutionizing for 6 hours at 540$^{\circ}$C and ageing for 3 hours at 155$^{\circ}$C. While tensile properties were not reported by Cheng et al., according to the data presented by Tiryakio{\u{g}}lu et al. (Fig. \ref{fig:Macro_v_micro}), this represents a slight decrease in yield strength. This is incongruent with measurements reported by Mori et al. and Zhao et al., which indicate that deformation improves mechanical properties.


For the most part, these studies show that mechanical properties of cast aluminum alloys are improved significantly by rotary deformation, however, they lack any insight into the cause of the change in mechanical properties. While the constitutive behaviour of this alloy in the as-cast (AC) condition has been characterized \cite{Roy.12a}, the effect of holding the AC structure at elevated temperatures for forming purposes, followed by deformation has unknown implications on the final heat treatment. The purpose of the this study is to characterize the modification of the microstructure of A356 from the AC condition through spinning (at varying intensities), followed by heat treatment. This is accomplished through microstructural observations on specimens with various thermomechanical histories, coinciding with extensive macro and microhardness measurements.

\section{Background}
\subsection{As-cast structure}
\begin{figure}[]
\centering
  \includegraphics[width=0.5\linewidth]{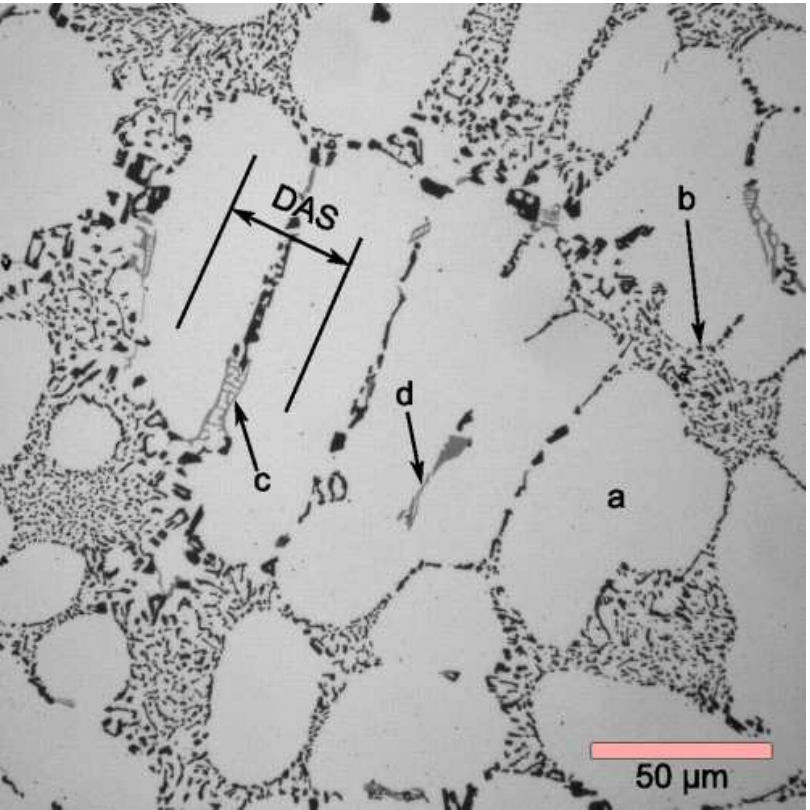}
\caption[]{A356 microstructure in the AC condition displaying DAS, (a) $\upalpha$-Al, (b) Al-Si eutectic, (c) intermetallic and (d) a secondary Mg-Si rich region.}
  \label{fig:MS}
\end{figure}

Hypoeutectic Al-Si-Mg alloys have an AC microstructure consisting of primary aluminum dendrites ($\upalpha$-Al) which form during the initial solidification phase, surrounded by an Al-Si eutectic phase (Fig. \ref{fig:MS}). Intermetallics may be present due to melt impurities, such as Fe which forms Al-Fe-Mg-Si structures. A356 is rarely employed in the AC condition owing to the coarse morphologies of eutectic-Si particles and non-uniform distributions of precipitates. Strength and elongation can be improved through heat treatment and chemical modification, with the latter achieved through the addition of small amounts of Na and Sr. These additions change the morphology of the eutectic-Si, rendering a structure which is less acicular and more fibrous and refined. Tempers applied to these alloys serve to modify the eutectic structure and refine/redistribute the Mg$_2$Si particles. The T6 treatment optimizes strength and ductility and is one of the more commercially common heat treatments for modified Al-7Si-0.3Mg (A356) alloy. The steps in the T6 process are: i) solution treatment at 540$^{\circ}$C for 4-12 hours, ii) quenching in water between 65-100$^{\circ}$ C, and iii) artificial ageing (precipitation treatment) at 155$^{\circ}$C for 2-5 hours according to ASTM B917/B917M-12. Ambient temperature ageing processes occurring between quenching and artificial ageing (natural ageing) should be minimized as it reduces the precipitation driving force necessary for artificial ageing \cite{Moller.07}.

As-cast microstructure refinement is achieved primarily through decreasing solidification time. This is readily observable in the dendrite arm spacing (DAS), which is primarily dependent on the cooling rate\cite{Flemings.74} and to a lesser degree, composition \cite{Easton.11}. Al-Si alloys have been found to have mechanical properties better correlated to DAS as opposed to grain size \cite{Goulart.06}. The distribution and morphology of eutectic-Si has a greater influence on mechanical properties than DAS \cite{Apelian.89}, with eutectic-Si particles having a scale approximately an order of magnitude less that the dendritic spacing. The consistency (i.e. uniform shape, size and distribution) of the precipitation hardening particles (Mg$_2$Si) occurring throughout the microstructure also dictates the mechanical properties.

McQueen et al. \cite{McQueen.98} conducted a comparative deformation study of A356 and SiC-A356 Metal Matrix Composite (MMC) in the AC condition at elevated temperatures and strain rates. It was posited that the rate-dependency of A356 flow stress at elevated temperatures is largely due to dynamic recovery as opposed to recrystallization in A356, the majority of which occurs in the eutectic. While recrystallization may not be disallowed, it likely does not play a significant role; McQueen et al. highlighted that this is particularly true in the $\upalpha$-Al phase as there is little solute to provide nucleation sites for new grains.

\subsection{Heat treatment}
The solution treatment is applied to induce three phenomena to occur: i) dissolution of Mg$_2$Si particles, ii) chemical homogenization, and ii) eutectic-Si fragmentation and spheroidization. The Mg$_2$Si precipitate that forms during the last stages of solidification is readily soluble in $\upalpha$-Al at the typical solutionizing temperatures \cite{Belov.05}, and will dissolve given enough time. In order to maximize the amount of Mg and Si in solution, a solutionizing temperature as close as possible to the equilibrium eutectic temperature is desirable. A temperature of 540$^{\circ}$C is high enough such that incipient melting at the grain boundaries is avoided. A356 has been shown to be completely homogenized after between 30-180 minutes \cite{Closset.86,Colley.11}, with longer times favoured to dissolve intermetallics \cite{Gustafsson.86,Wang.01-solidification}. Too long a solutionizing treatment can cause suboptimal eutectic particle sizing through coarsening \cite{Apelian.89,Shivkumar.90}.

After solutionizing, quenching suppresses precipitation to maximize the degree of supersaturation at the start of artificial ageing \cite{Zhang.96,Sjolander.10}. Artificial ageing should take place immediately after quenching to minimize any ageing at ambient conditions (natural ageing).  Edwards et al. \cite{Edwards.98} identified that the particle morphology coinciding with peak artificial ageing is a rod shaped precipitate, having a nanometer scale needle-like structure. Further ageing generates equilibrium Mg$_2$Si platelets or metastable rods and a corresponding decrease in strength associated with over-ageing. Colley \cite{Colley.11} found that the peak aged conditions was reached after approximately 1 hour at 200$^{\circ}$C, 3 hours at 180$^{\circ}$C or 8 hours at 150$^{\circ}$C for A356 when artificially aged immediately after quenching.

\subsection{Implications}
The mechanical properties of heat treatable aluminum casting alloys are dependant on microstructural features spanning several length scales, which are all affected by rotary forming at elevated temperatures. There is the potential for mechanical processing to affect the microstructure, and due to the AC material being held at temperature prior to forming, there is also potential for thermal effects. This includes changes to the precipitate structure and distribution, in addition to the final eutectic-Si particle distribution in the T6 condition.

\section{Material and experiment methodology}\label{sec:Material}
The material investigated in this study is strontium-modified, low pressure die-cast (LPDC) A356 supplied by a North American aluminum alloy wheel manufacturer with the nominal chemical composition given in Table \ref{table:composition}. This was supplied in the form of LPDC wheels, which were then machined to form blanks with a $\sim$330 mm diameter and $\sim$10 mm thick to be compatible with an experimental rotary forming apparatus. Forming experiments took place approximately 45 days after casting.

\begin{table}
    \centering
    \caption{A356 composition in wt-\%.\label{table:composition}}
    \begin{tabular}{lllllll}
    \toprule
    Element & Si & Mg & Fe & Na & Sr & Al\\
    \midrule
    Range (wt-\%) & 7.04 & 0.39 & 0.13 & $\sim$0.002 & $\sim$0.005 & Balance\\
    \bottomrule
    \end{tabular}

\end{table}

\begin{figure}[]
\centering
  \includegraphics[width=0.9095\linewidth]{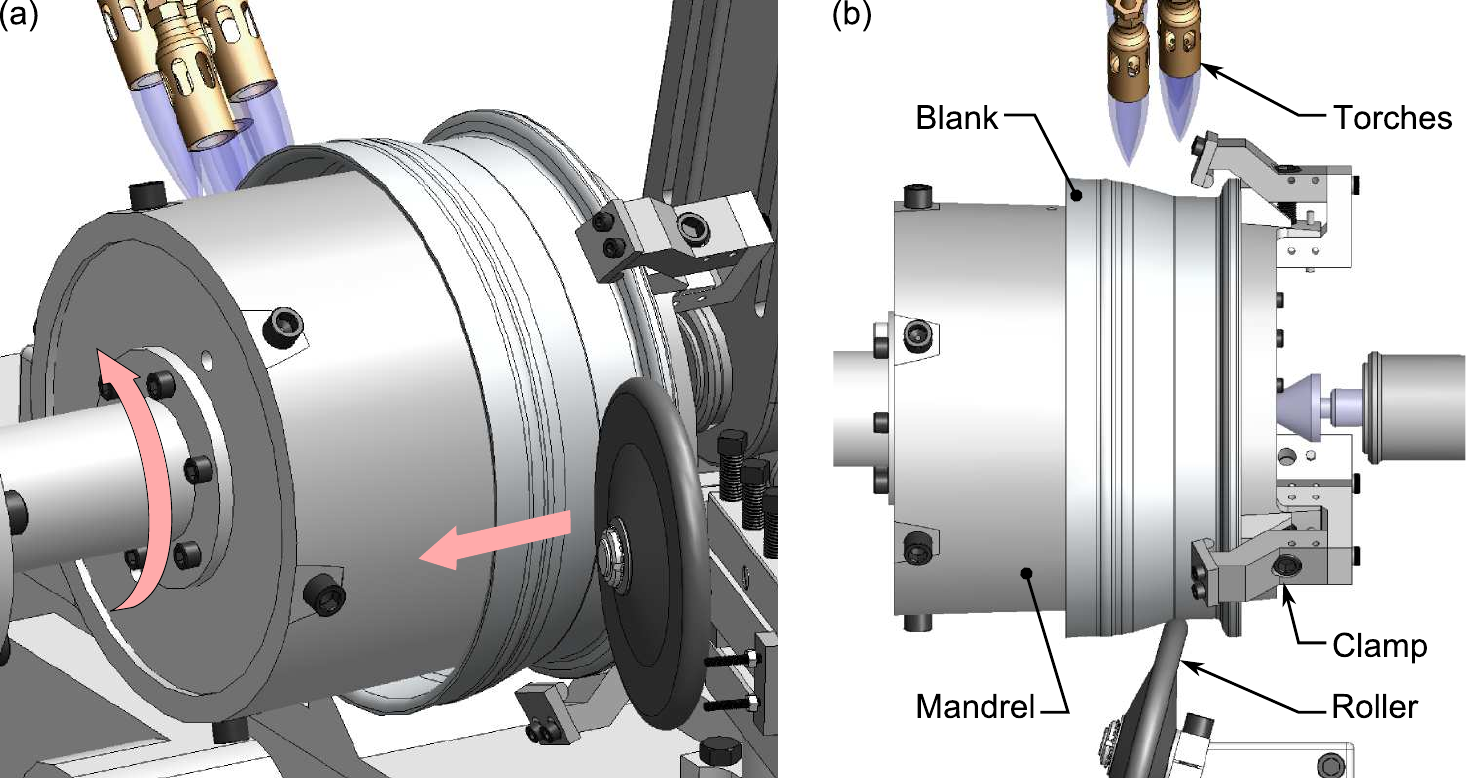}
  \caption[]{3D depiction of the experimental rotary forming apparatus and forming directions in (a), top-down view of specific apparatus components in (b).}
  \label{fig:Process}
\end{figure}

This apparatus consisted of a customized manual lathe with a tapered mandrel mounted directly to the spindle, supported by a tailstock with a live center. The mandrel was fitted with a manually-actuated clamping apparatus designed to accommodate the blank at both ambient and elevated temperatures. The toolholder of the lathe was converted to hold a roller with an overall diameter of 120 mm and a nose radius of 10 mm at an attack angle of 15$^\circ$. This arrangement is shown in Fig. \ref{fig:Process}.

Propane torches with a total heat output of 82 kW were employed to uniformly heat the blank from ambient temperature to a target temperature of 375$^\circ$C by rotating the blank at 20 RPM through the influence of the torches. This preheat temperature was selected as it is where the constitutive behaviour of the material transitions to being rate dependent and exhibits little to no work hardening \cite{Roy.12a}. Heat transfer to the mandrel was abated by the application of a refractory-type coating\footnote{Foseco DYCOTE 32} to the inner diameter of each blank. The time required to preheat the blank to the forming temperature varied from 17 to 23 minutes, as heating was periodically interrupted to adjust the clamping mechanism to account for thermal expansion. Blanks having a better mandrel fitment took longer to heat up because of increased heat transfer to the mandrel. The length of time each sample was exposed to elevated temperatures is given in Table \ref{table:HeatTimes}. The uniformity of heating was verified using a single blank instrumented with 3 embedded type-K thermocouples spaced axially equidistant and 30$^\circ$ circumferentially and recorded with a wireless data acquisition (DAQ) system. The circumferential temperature difference was negligible, and the axial temperature distribution was within 8$^\circ$C.

\begin{table}
\centering
\caption{Heating and forming times for each sample with mean $H_{\mathrm{V}5}$ in found after forming. \label{table:HeatTimes}}
\begin{tabular}{lllll}
\toprule
\multirow{2}{*}{Sample} & Heating time & Forming time & Total Processing time & Mean $H_{\mathrm{V}5}$ \\
                        &(min.)        &(min.)        & (min.)                  & kg/mm$^2$ \\
\midrule
L & 21.2 & 1.0 & 22.2 & 53.00 \\
M & 17.3 & 1.4 & 18.7 & 55.53 \\
H & 22.9 & 2.5 & 25.4 & 49.27 \\
\bottomrule
\end{tabular}
\end{table}

Non-contact blank surface measurements were attempted using infrared thermocouples as per Mori et al. \cite{Mori.09} using two different sensors\footnote{Exergen IRt/c.1X-K-440F/220C and IRt/c.10A}, however, it was found that the low emissivity of the blank material and surface irregularities precluded accurate measurements as compared to a contact method. Therefore, axial surface measurements of the temperature of the blank were performed manually with a type-K thermocouple surface probe\footnote{Omega model number 88108} on each experiment every 3 minutes, as well as immediately before and after forming. During the course of forming, the surface temperature along the axis of the workpiece remained above 342$^\circ$C.

Once the blank was at the appropriate temperature, the mandrel speed was increased to 281 RPM and the roller was brought into contact with the blank with a radial speed of approximately 30 mm per minute. A thread-cutting feed screw was then engaged to move the roller axially at a rate of 0.21 mm per revolution while continuing to heat with the torches. Once the forming pass was complete, the clamps and blank were removed from the mandrel and left to air cool, avoiding potential distortion from quenching. This was repeated to produce three workpieces with increasing levels of deformation. The workpiece with the least deformation corresponds to that normally seen in spinning operations, with the others corresponding to increasing levels of `overspinning', with the peak-formed specimen approaching deformation conditions experienced in flow forming \cite{Wong.03}. These workpieces are further referred to as `L', `M' and `H' denoting low, medium and high levels of deformation.

\begin{figure}[]
\centering
  \includegraphics[width=0.5\linewidth]{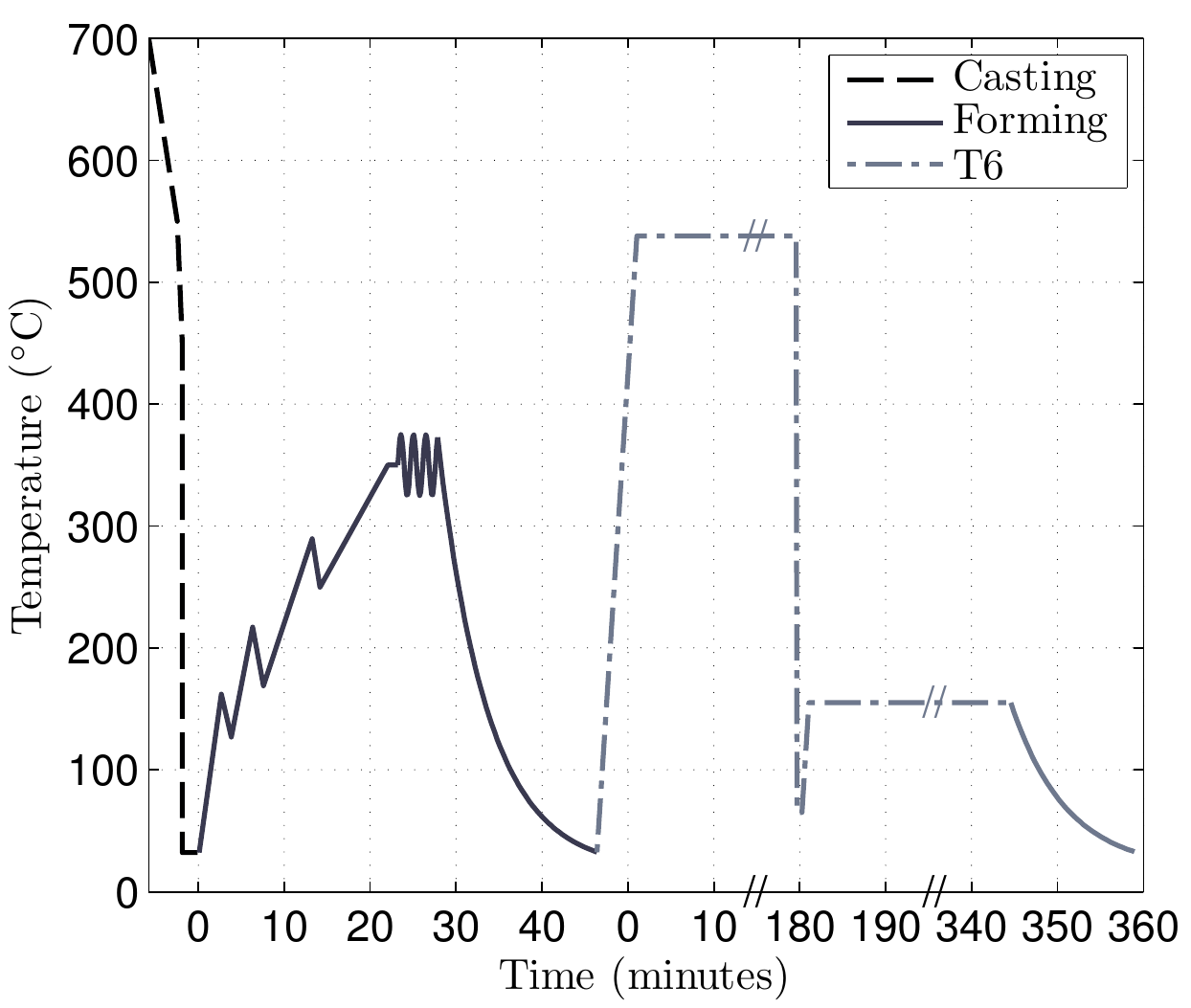}
  \caption[]{Temperature-time profile for casting, forming and final T6 heat treatment.}
  \label{fig:TT}
\end{figure}

Once the workpieces had cooled, sections were removed for analysis and the remainder of the material was subjected to a T6 heat treatment within 48 hours of forming. Along with material from an unformed blank, the formed materials were solutionized at 538$^\circ$C for 3 hours, quenched in water at 65$^\circ$C and immediately artificially aged at 155$^\circ$C for 3 hours. A complete representative thermal history is given in Fig. \ref{fig:TT} for all processing steps. Smaller samples were extracted from an unformed blank in the AC condition and the corresponding region in the H sample. These smaller AC samples were the subject of an ageing study and a number were deep-etched to reveal eutectic particle morphology as presented in Section \ref{sec:Ageing}. In conjunction with those extracted from the H workpiece, these samples were then subjected to microhardness and SEM analyses as discussed in later sections.

Various cross-sections were extracted from an unformed blank and the formed blanks. Cross-sections in the AC, formed, and T6 (AC-T6 and formed-T6) states were then mounted and polished with alumina to at least 1 $\upmu$m.  The cross-section profiles were then optically digitized\footnote{Employing a Hewlett Packard ScanJet 4200C}. Hardness profiles were generated via a custom, numerically controlled stage with a resolution of 20 $\upmu$m installed on a Vickers-Armstrong macrohardness tester. Each indentation site was imaged using a digital single-lens reflex camera\footnote{Canon EOS Rebel T2i fitted with a Martin Microscope MM-SLR} adapted to the hardness tester's microscope. In the case of the formed samples, hardness measurements were performed within 72 hours of forming. A lesser number of $H_{\text{V}0.01}$ measurements were performed with a Buehler Micromet II microhardness tester. All hardness measurements presented in this study conform to ASTM E384.

Optical image processing was conducted via the MATLAB\footnote{MATLAB is a trademark of The MathWorks Inc., Natick, MA} Image Processing Toolbox. Energy-dispersive X-ray (EDX) spectroscopy analysis and Si particle imaging was conducted with an Hitachi S-3000 electron microscope in backscatter electron mode with an accelerating voltage of 7keV.

\section{Microstructure and hardness}
\begin{figure}[]
\centering
  \includegraphics[width=0.5\linewidth]{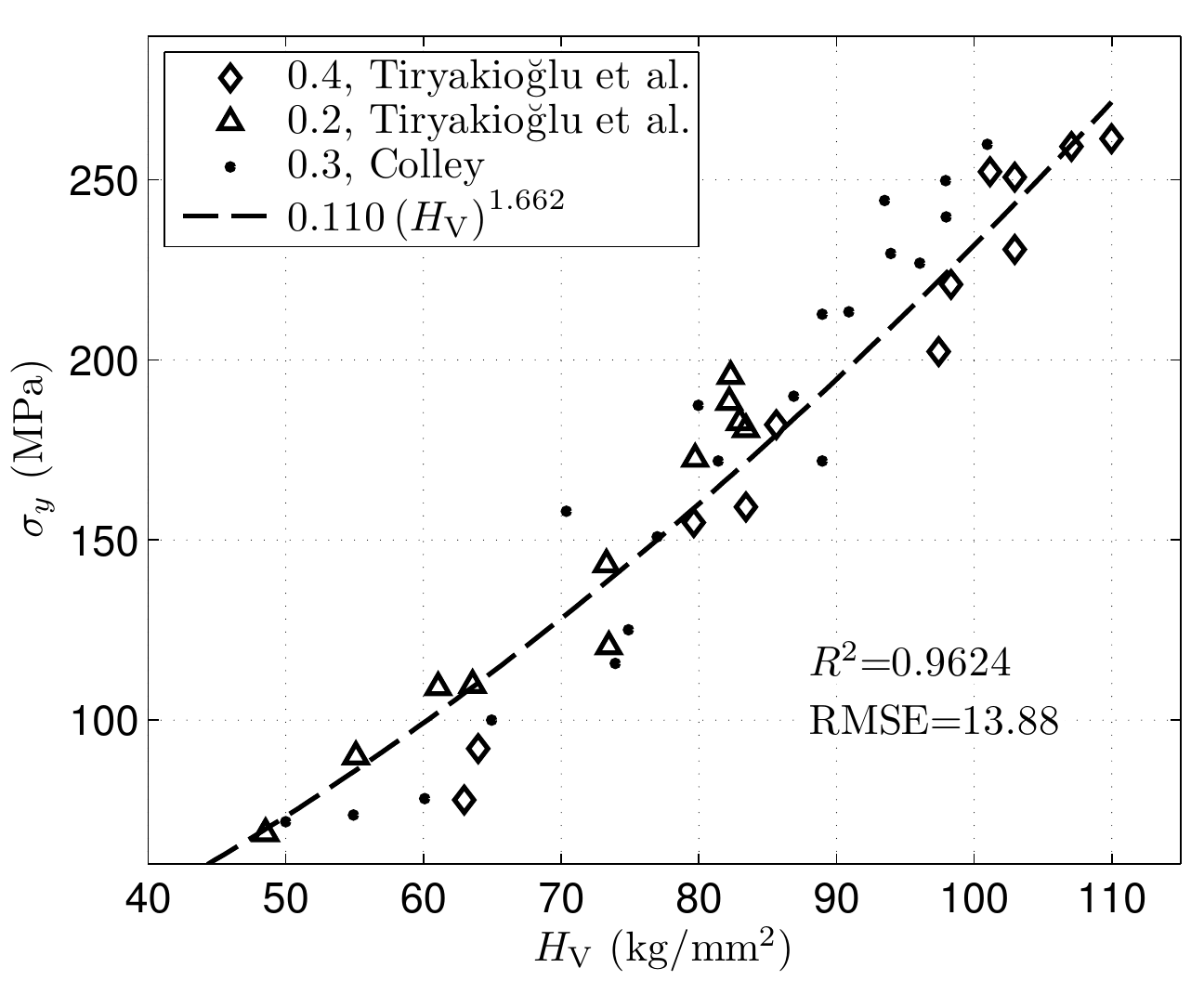}
  \caption[]{$\sigma_y$ vs. $H_{\text{V}}$ relationship developed from $\sigma_y$ vs. $H_{\text{RF}}$ measurements made by Tiryakio{\u{g}}lu et al. [28] on underaged Al-7\%wtSi-x.x\%wt-Mg. Data from Colley [20] for over and underaged Al-7\%wt-Si-0.3\%wt-Mg plotted for comparison. Fitting based on underaged results.}
  \label{fig:Macro_v_micro}
\end{figure}

As pyramidal diamond hardness measurements are directly proportional to flow stress \cite{Tabor.51}, with appropriate correlation, hardness values have been used for some materials to infer local yield strength. For heterogeneous foundry alloys such as A356, local anisotropy due to casting parameters may preclude accurate assessments of local yield strength even for large indents at the macro scale. Bulk yield strength has been previously correlated to macrohardness measurements for this type of material. Fig. \ref{fig:Macro_v_micro} shows selected bulk yield strength versus hardness results reported by Tiryakio{\u{g}}lu et al. \cite{Tiryakioglu.03} for underaged A356 with low and high levels of Mg, converted to $H_{\mathrm{V}}$ (original data was reported as $H_{\mathrm{RF}}$). A non-linear least-squares fit of this data suggests that $\sigma_y= f\left(H_{\mathrm{V}} \right)$ is best described by a power-law relationship. Adding similar data from Colley \cite{Colley.11} for A356 (with a mean DAS of 30 $\upmu$m) in both the over and underaged conditions shows good agreement, particularly at lower hardness values. The overall goodness of fit of the power-law relationship is greater than 0.95, and root mean square error (MSE) is approximately 14 MPa over all experimental data. In the present work, this correlation is used to infer relative changes in bulk strength due to processing.

\begin{figure}[]
\centering
  \includegraphics[width=\linewidth]{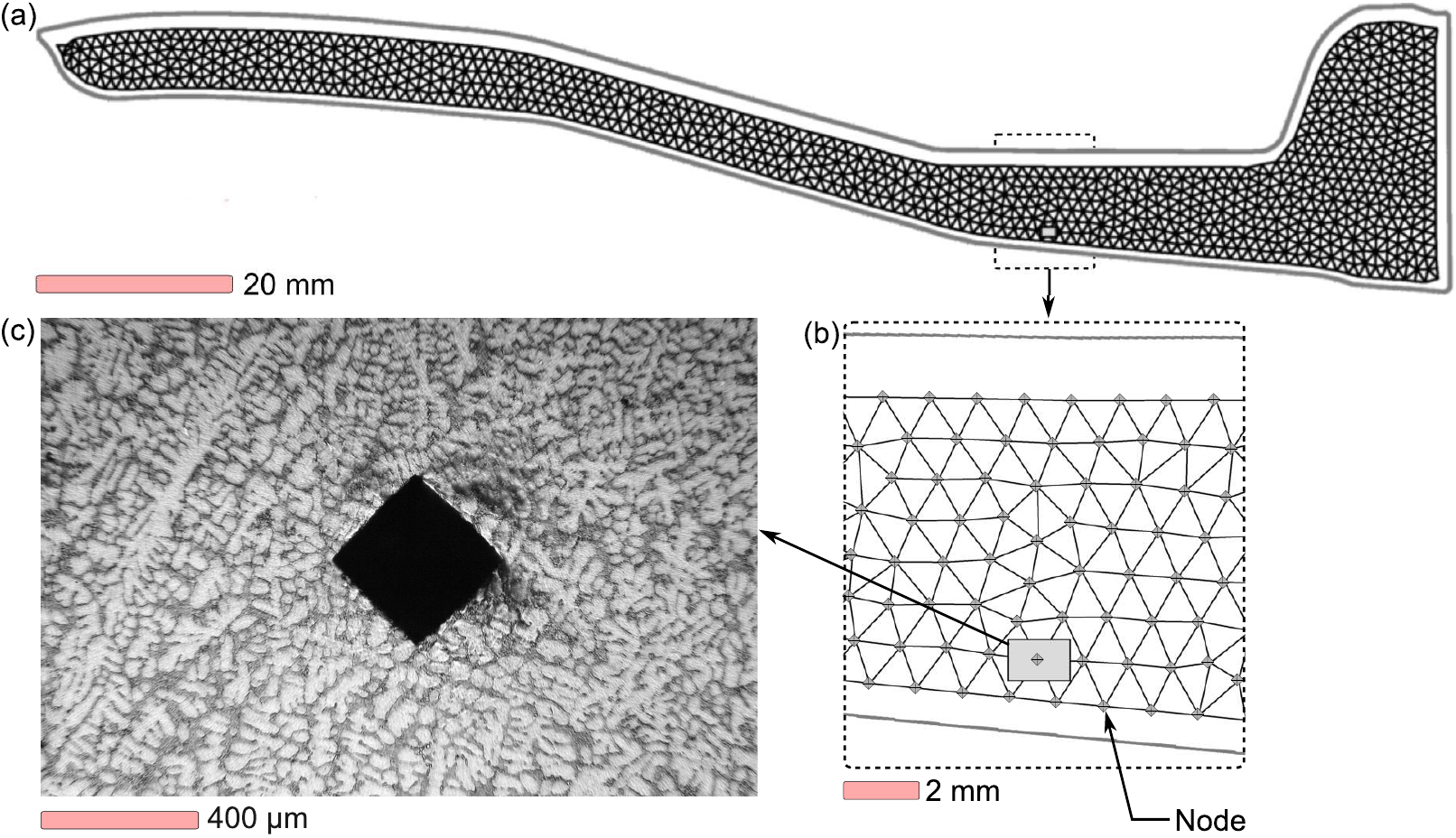}
  \caption[]{Hardness profile mesh detail. The dashed lines in (a) enclose a typical area shown in (b), which shows a typical low resolution micrograph field defined at each node in (c).}
  \label{fig:Phase}
\end{figure}

In total, 9 blank cross-sections were analyzed through hardness profiling. These were axial and circumferential cross-sections of the as-cast blank, axial cross-sections of as-formed blanks and all axial cross-sections in the T6 condition. Each digitized cross-section was scaled to ensure indents were spaced far enough from the specimen edge, and then uniformly meshed using Delaunay triangulation. The algorithm employed created a uniform mesh with a minimum distance between nodes of 450 $\upmu$m, with between 950-1100 nodes per profile. An example of the mesh produced for specimen L is shown in Fig. \ref{fig:Phase}. Indentations were then placed and imaged at the location of each node with the aid of a bespoke computer-numerically-controlled stage. Imaging each indentation site permitted automatic measurement of indentation size through digital image processing. Comparing manually measured indentations of a cross-section with those processed automatically, results obtained with image processing agreed with manually measured indents within 2\%.

\subsection{As-cast and as-formed materials}
\begin{figure}[]
\centering
  \includegraphics[width=\linewidth]{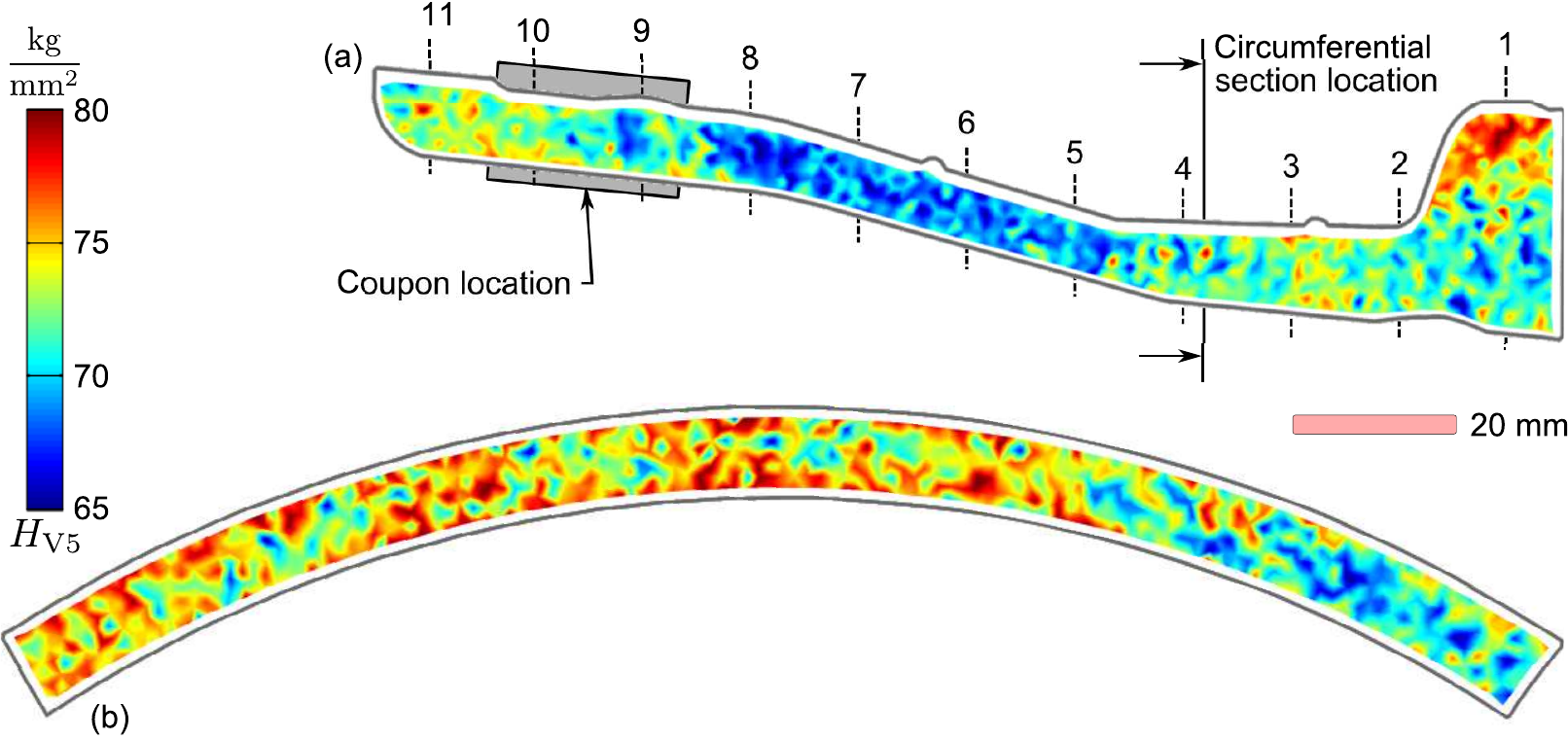}
  \caption[]{Hardness profile of the as-cast blank sectioned axially (a) and circumferentially (b) at plane indicated. Shaded region indicates location of samples for thermal treatment study. Numbered dashed lines indicate DAS measurement locations.}
  \label{fig:AC_mont}
\end{figure}

Fig. \ref{fig:AC_mont} shows the results of hardness measurements performed on axial and circumferential (72$^\circ$) profiles from the as-cast blank. Also shown are 11 axially equidistant section markers to assist in tracking changes through processing. Hardness measurements performed on the two profiles showed a similar range of hardness. Axially, the highest hardness was found to be at either end of the sample, with the center being the softest. Circumferentially, a gradient in hardness was observed. The variations in hardness in both directions is attributed to differences in DAS and eutectic phase fractions caused by variations in solidification time and the transport of Si-enriched liquid during solidification.

\begin{figure}[]
\centering
  \includegraphics[width=\linewidth]{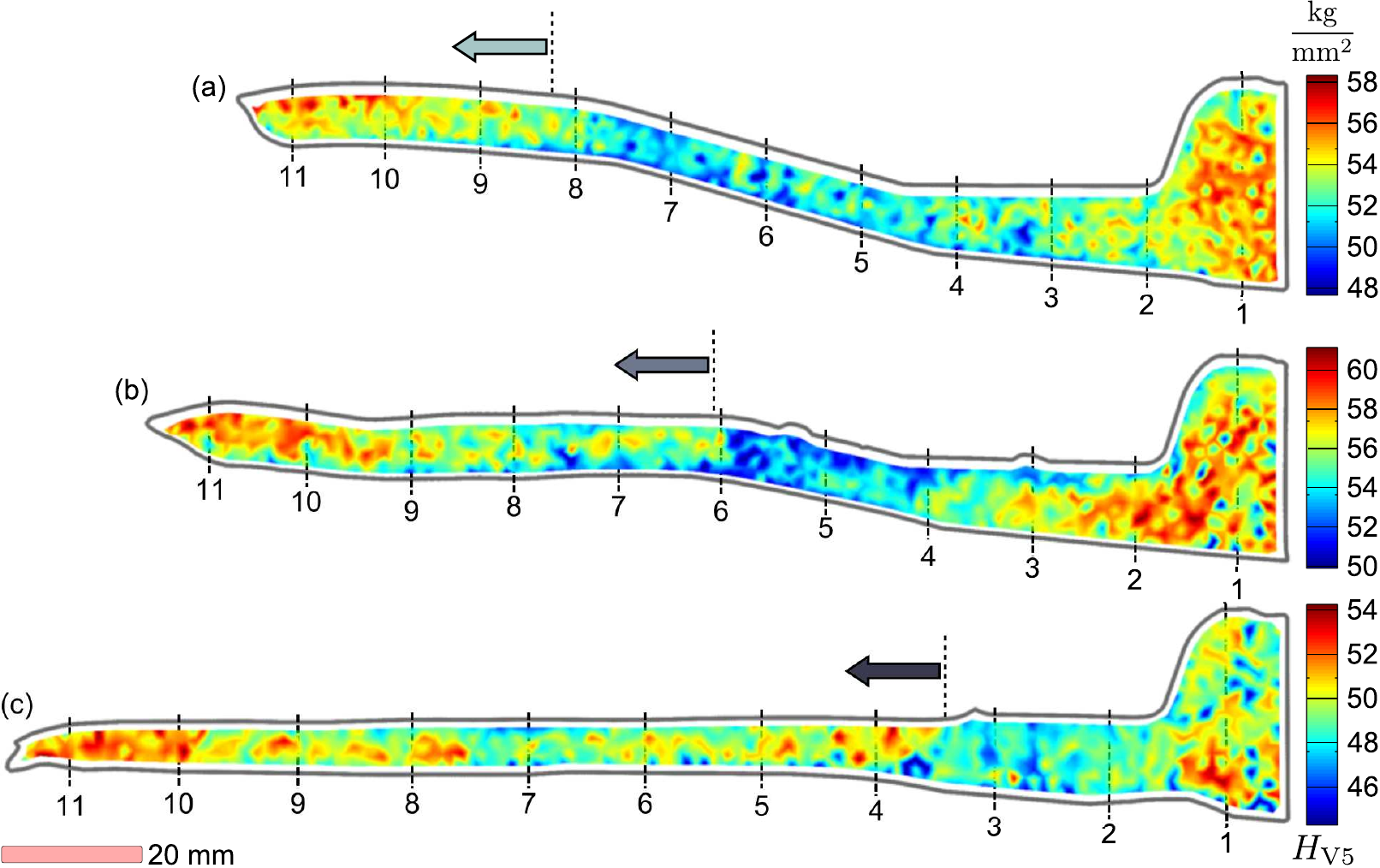}
  \caption[]{Comparative results of as-formed sections with least deformation (a), mid-formed (b) and peak-formed (c). Arrows indicate forming start point, and dashed lines indicate DAS measurement locations.}
  \label{fig:All_Pre}
\end{figure}

The hardness profiles of each of the as-formed sections, shown in Fig. \ref{fig:All_Pre}, demonstrate the large changes that occur from the AC condition. In all cases, the mean hardness has dropped significantly, in-line with the relative time at temperature each sample underwent. Fig. \ref{fig:All_Pre} has independent contour levels centered about the mean hardness (Table \ref{table:HeatTimes}), and has been annotated with arrows identifying the point of initial roller contact in order to delineate formed regions. In the unformed regions, the hardness distributions show similar trends to the AC condition albeit with a reduction in average hardness. A similar effect is observed for the high hardness locations at the tips of the specimens (locations 10 and 11) in the formed regions. The remaining positions in the formed regions show increased hardness values relative to the peak hardness in the sample when compared with the AC condition. Additionally, the increased hardness values in these areas (sections 8 and 9, 6 - 9, and 4 - 9, respectively), are higher closer to the surfaces deformed by the roller. This type of localized distribution has also been seen in the rotary forming of steel \cite{Roy.09}.

\begin{figure}[]
\centering
  \includegraphics[width=\linewidth]{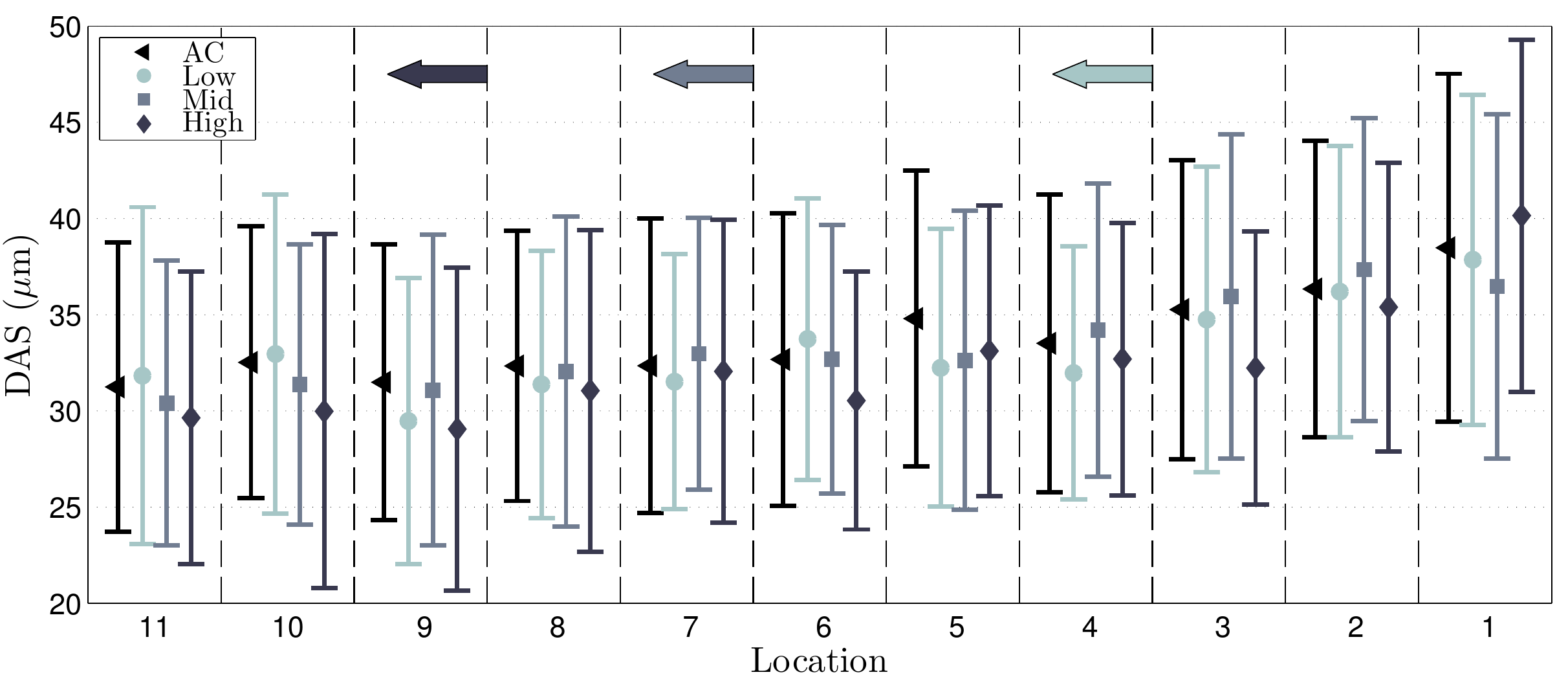}
  \caption[]{Mean DAS measurements of unformed and formed blanks. Arrows indicate axial start of forming for each specimen. Error bars indicate $\pm$ 1 standard deviation.}
  \label{fig:DAS}
\end{figure}

In order to track the effects of processing on DAS, five optical micrographs/fields were selected at random along each of 11 locations shown in Figs. \ref{fig:AC_mont} and \ref{fig:All_Pre} at a depth between 2--3 mm from the outer diameter. Approximately 300 discrete measurements were performed across each of the 5 fields per section. The results of this analysis, shown in Fig. \ref{fig:DAS}, indicate that the DAS increases with cross-section thickness in the AC material. As larger DAS should correspond to lower hardness values, the cause for elevated hardness in locations 1--3 for all specimens is presumed to be related to the presence of elevated levels of eutectic. Comparing results in each condition at each section, these results suggest that the DAS has not been significantly affected by the deformation levels achieved during forming. This would also indicate that DAS is not an effective quantifier of highly local changes to microstructure imposed by rotary forming processes, particularly when high levels of deformation has been shown to decimate dendritic structure in other studies \cite{Zhao.11}.

\subsection{Heat treated materials}
\begin{figure}[]
\centering
  \includegraphics[width=.9224\linewidth]{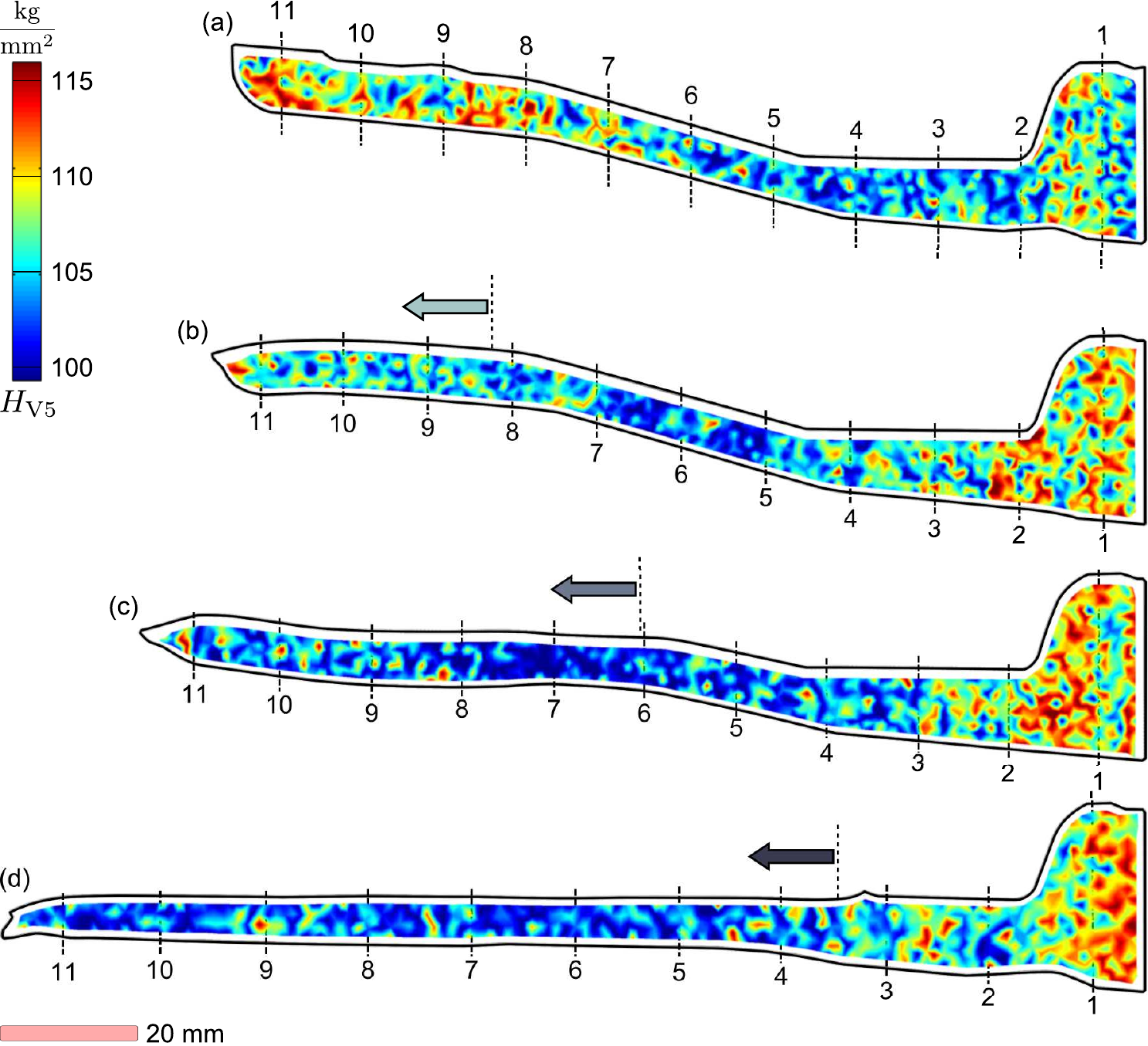}
  \caption[]{Comparative results of sections in the T6 condition with no deformation (a) versus increasing levels of deformation (b-d). Arrows indicate forming start point.}
  \label{fig:All_T6}
\end{figure}

The formed specimens in the T6 condition show an overall increase in hardness (Fig. \ref{fig:All_T6}), however the hardness distribution in the specimens changed significantly. Here, formed regions show decreasing hardness with increased deformation. The elevated hardness region found in the unformed workpiece (Fig. \ref{fig:All_T6}a, location 11), $\sim$ 115 kg/mm$^2$, is progressively reduced to approach a nominal hardness value of approximately 100 kg/mm$^2$ in the H specimen. Unformed regions in specimens subjected to rotary forming (Fig. \ref{fig:All_T6}b-c, location 1) are found to be marginally higher than the unformed workpiece by approximately 5 kg/mm$^2$ (Fig. \ref{fig:All_T6}a, section 1). The 15 kg/mm$^2$ difference in hardness between unformed and heavily deformed material (referring to sections 10 and 11) represents a 20\% decrease in strength (Fig. 4).

Clearly, the microstructure has been significantly altered by the forming process. Even though the DAS was measurably diminished as levels of deformation increased, this did not translate to increased hardness. Furthermore, while the hardness measurements decreased at locations that have experienced both thermal and mechanical processing, the hardness marginally increased in regions where only thermal effects occurred.

\section{Processing effects on microstructure}\label{sec:Ageing}
In order to ascertain the effects of holding the AC material at an elevated temperature before forming, coupons (location and size given on Fig. \ref{fig:AC_mont}) were extracted from an AC blank and held at elevated temperatures in a nitrate salt bath (60\% potassium nitrate, 40\% sodium nitrate) for varying lengths of time. Samples were left to air cool upon removal from the salt bath. Hardness measurements were made on each sample before the treatment and within 30 minutes of cooling to ambient temperature. The microstructure of select samples was also assessed using optical and electron microscopy. This work was aimed at determining the effects of holding the material at an elevated temperature (or ageing) prior to forming.

The temperatures selected for this work were based on potential forming temperatures and include: 300, 350, and 400$^\circ$C, as well as, the solutionizing temperature of 540$^\circ$C. Target hold times, selected to span the potential breadth of forming operations, were: 2, 10, 20 and 50 minutes. The temperature history of each sample was monitored with a thermocouple. The approximate time to cool to 100$^\circ$C for all specimens was 3.5 minutes.

\subsection{Thermal effects on hardness}
\begin{figure}[]
  \centering
\includegraphics[width=0.5\linewidth]{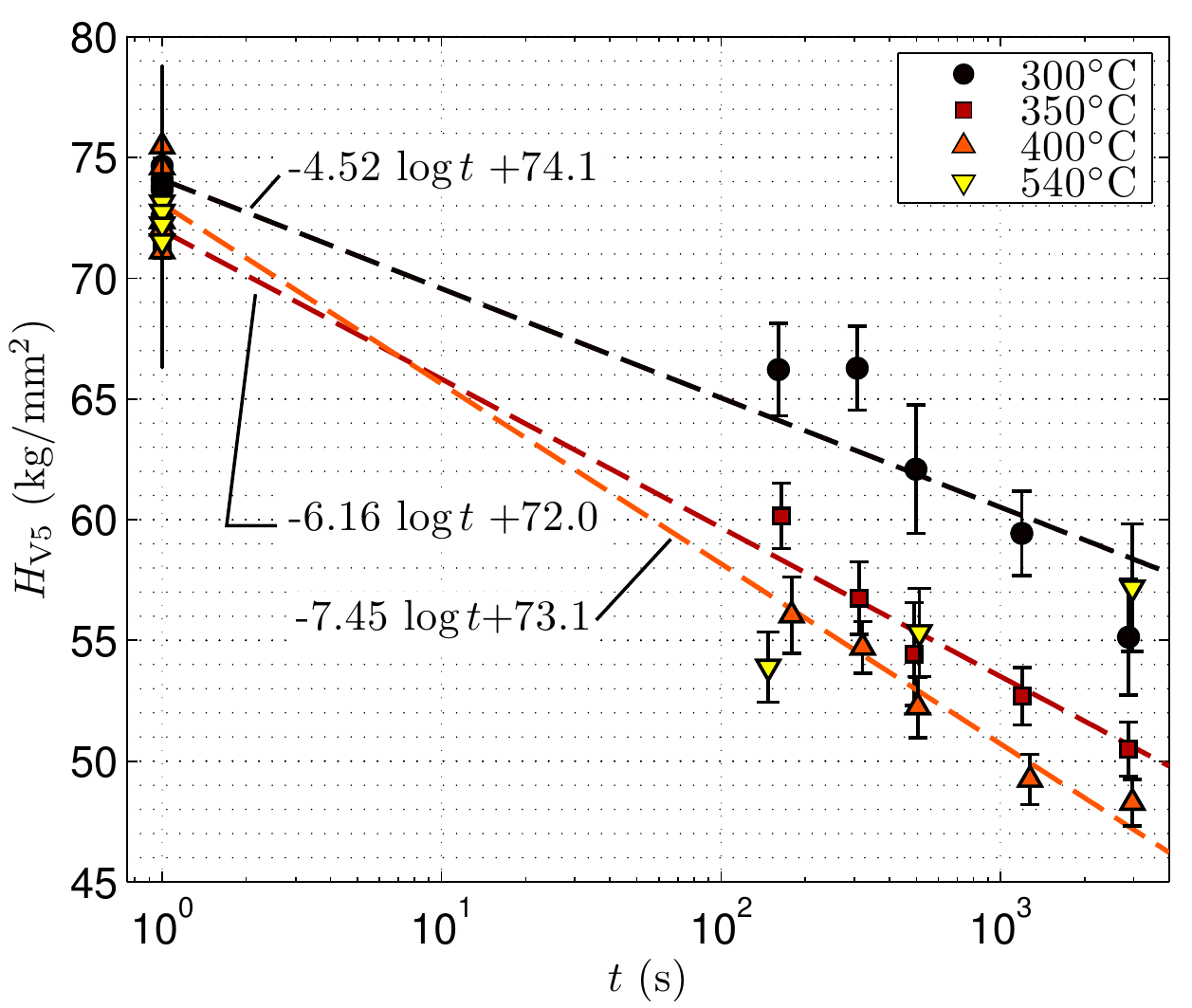}
  \caption[]{Comparative $H_{\mathrm{V}5}$ results of specimens initially in the as-cast condition and after holding at various temperatures and times.}
  \label{fig:HomoMacro}
\end{figure}

\renewcommand\thelinenumber{\color{black}\arabic{linenumber}}
The average hardness and standard deviation for each hold temperature are plotted as a function of hold time in Fig. \ref{fig:HomoMacro}. For all temperatures below 540$^\circ$C, there is a clear power law drop in hardness versus hold time, with better agreement at 350 and 400$^\circ$C. Furthermore, as temperature increases, the standard deviation in the hardness measurements diminishes. Apelian et al. \cite{Apelian.89} reported that the equilibrium solubility of Mg and Si in solid aluminum increases by an order of magnitude when the temperature is increased from 300 to 400$^\circ$C. Thus, increasing the temperature will affect the following phenomena:
\begin{itemize}
\item Coarsening of Mg$_2$Si precipitates that formed during initial casting;
\item Complete or partial dissolution of small Mg$_2$Si precipitates;
\item Eutectic-Si spheroidization beyond initial fragmentation; and
\item Eutectic-Si coarsening beyond spheroidization.
\end{itemize}

The hold temperatures employed for the experiments summarized by Fig. \ref{fig:HomoMacro} are characteristic of those that material may encounter during forming. The net effects of microstructural changes on hardness can be expressed as a function of time $t$ and temperature $T$:
\begin{linenomath*}
\begin{equation}
\Delta H_v=\left(-2.93\times 10^{-2}T + 4.21\right)\log t
\end{equation}
\end{linenomath*}
and implies that there is no thermal effect on the microstructure ($\Delta H_v=0$) below 144$^\circ$C.

In the case of the 540$^\circ$C results, there is no power law drop in hardness versus time observed as with the other temperatures. Consistent with samples tested at other temperatures, there was a large initial drop observed in the specimen held for 2 minutes. However, the hardness increases from this point on. The effects of precipitation occurring during the slow cooling of the samples are superimposed. With longer temperature holding times, there is a progressive increase in precipitate dissolution, leading to higher levels in solution. The low air cooling rate and the potential for natural ageing results in increased hardness with time, coincident with increased levels of dissolution attained at temperature.
\subsection{Microstructure}
\begin{figure}[]
\centering
  \includegraphics[width=\linewidth]{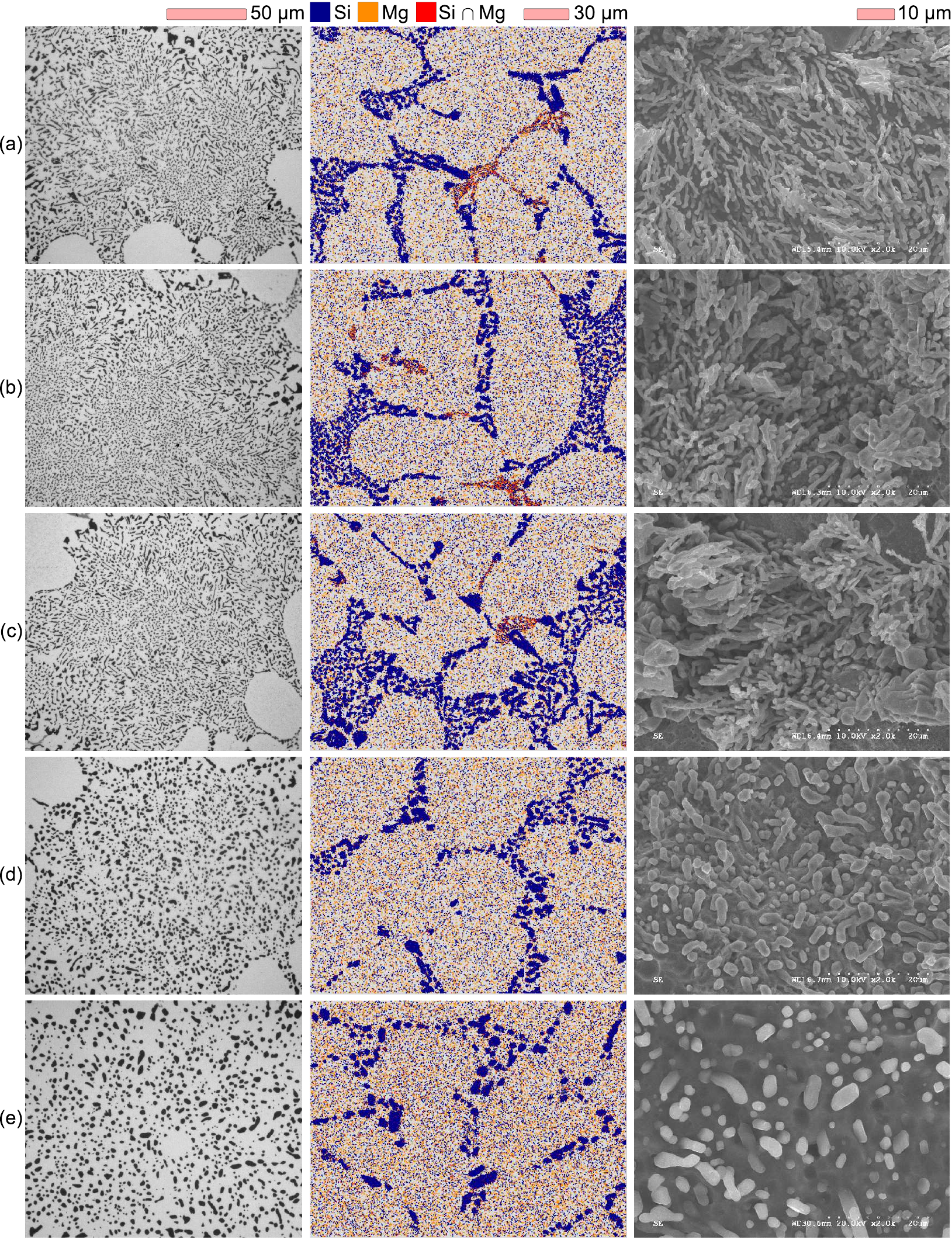}
  \caption[]{Optical microstructure images, EDX element maps and SEM images of eutectic particle morphologies following deep etching for specimens in the as-cast condition (a), held for 50 minutes at 300 (b), 400 (c), 540$^\circ$C (d) and the T6 condition (e).}
  \label{fig:montNorm}
\end{figure}

\begin{figure}[]
\centering
  \includegraphics[width=\linewidth]{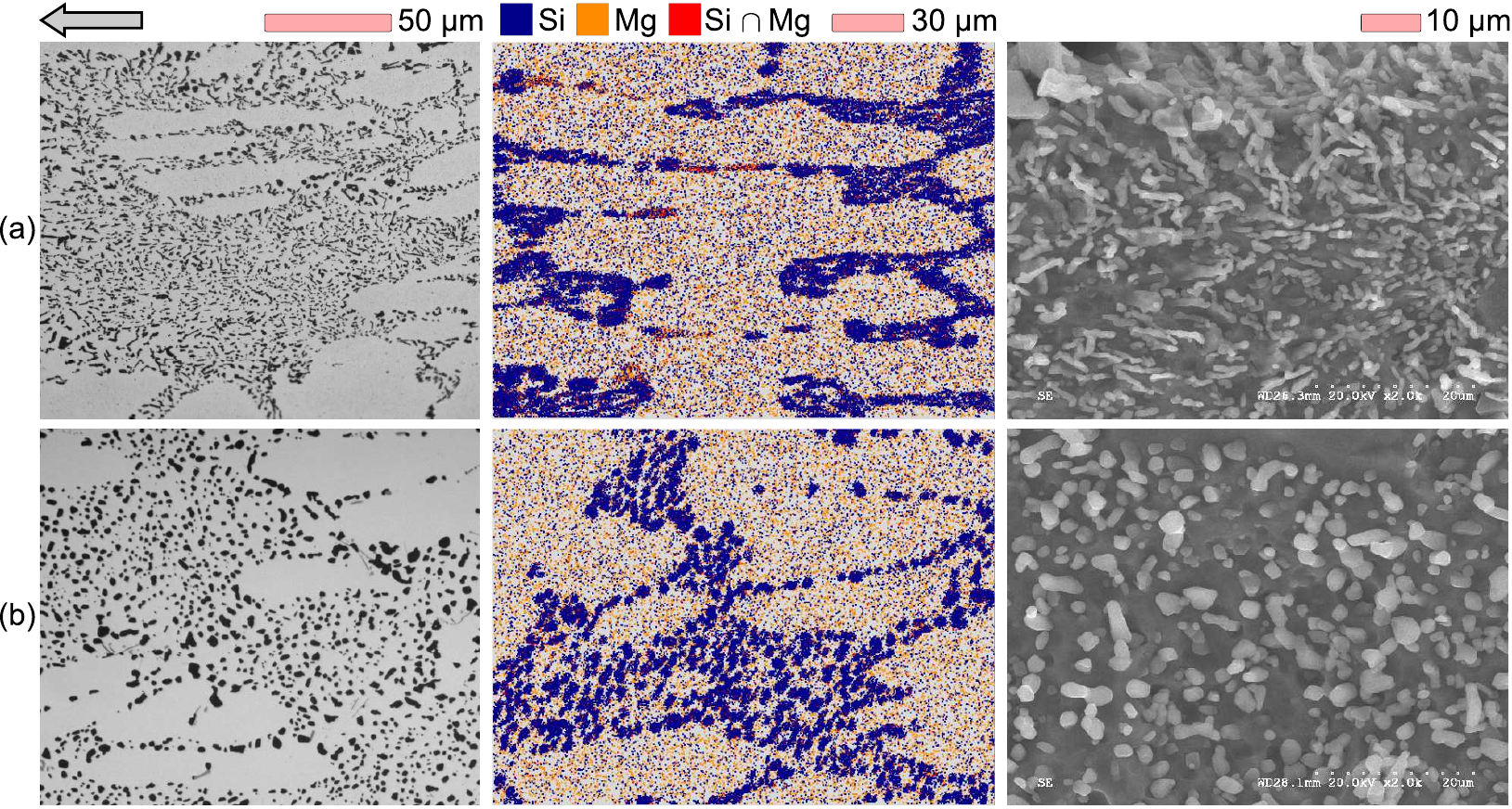}
  \caption[]{Optical microstructure images, EDX element maps and SEM images of eutectic particle morphologies following deep etching for specimens in the as-formed (a) and formed-T6 condition (b). Arrow indicates forming direction.}
  \label{fig:montFF}
\end{figure}

To examine the effects of hold temperature on the microstructure, the specimens held at each respective temperature for 50 minutes were analyzed via optical microscopy and EDX.  These specimens were also deep etched through immersion in Keller's etch (10\% hydrofluoric acid and 5\% hydrochloric acid by volume in water) for 50 minutes to reveal the insoluble silicon particles present in the eutectic. The results of this analysis, presented in Fig. \ref{fig:montNorm}, show how the eutectic-Si structure evolves with increased holding temperature. Subtle modification of the eutectic structure is evident from the optical microscopy, while the SEM images obtained after etching showing coarser features with increasing hold temperature. Specimens held at 300$^\circ$C show that larger eutectic-Si branches have been rounded and are joined by less refined fiber morphology. Increasing the temperature to 400$^\circ$C shows a continued evolution of this morphology resulting in fewer, thicker branches being observed. At the solutionizing temperature of 540$^\circ$C, the particles are fully fragmented and spheroidization is evident. The EDX results show that localized Mg-bearing structures are present up to 400$^\circ$C. These are expected to be Mg$_2$Si (outlined in red/orange in Figs. \ref{fig:montNorm} and \ref{fig:montFF}); however, they may also be intermetallics. Once the solutionizing temperature is reached, there is no evidence of these localized Mg-bearing structures present. While the EDX observations do not show the evolution in distribution of Mg at 300 and 400$^\circ$C from the AC state, the absence of regions containing concentrated Mg between holding at 400 and 540$^\circ$C are congruent with the observations made regarding the macrohardness results.

A sample of undeformed material following the complete T6 heat treatment was also analyzed using this methodology. The results of this analysis are also presented in Fig. \ref{fig:montNorm}.  The distribution of Mg in this sample is approximately the same compared to the sample held at 540$^\circ$C for 50 minutes. The eutectic-Si in the T6 sample has also spheroidized to a greater extent and some coarsening has occurred as characterized by larger and fewer particles with the same field size.

This methodology was further applied to analyze formed material before and after a T6 heat treatment. Specimens were extracted from a location approximately 1 mm from the roller interface in the sample that experienced the largest deformation. The axial location of the specimens coincided with the unformed specimens employed to evaluate the effect of hold temperature/time. The resulting micrographs for this material are shown in Fig. \ref{fig:montFF}.  Prior to heat treatment, the eutectic-Si particle size has decreased and has been compacted in line with the deformation. There is less evidence of spheroidization having occurred, as the Si morphology is observed to be small, short fibers/plates. The EDX maps suggest that the Mg-bearing structures have consolidated on the edges of the dendrite arms, appearing as plates oriented parallel to the forming direction. The morphology and distribution of these structures explains the hardness profiles seen in the spun material prior to heat treatment (Fig. \ref{fig:All_Pre}c), where regions of elevated macrohardness were found coinciding with formed regions. In the formed-T6 condition, localized Mg is absent as in the case of the unformed material, however the eutectic structure differs. While spheroidized, the eutectic particles are found to be appreciably smaller in count and size for equivalent field sizes than those observed in the unformed material.

\subsection{Eutectic particle shape and size}
The Lifshitz, Slyozov and Wagner (LSW) coarsening model \cite{Lifshitz.61,Wagner.61} provides a means of quantifying eutectic particle size evolution with time:
\begin{equation}\label{eq:LSW}
k=\frac{\bar{d}^{\;3}-\bar{d}_0^{\;3}}{t}
\end{equation}
Competitive coarsening driven by diffusion, as described by the LSW model \cite{Lifshitz.61,Wagner.61}, predicts a steady-state lognormal distribution of particle sizes about $\bar{d}$ once fragmentation is complete \cite{Tiryakioglu.08}. The presence of these hard Si particles within the softer Al matrix results in a strengthening effect due to the eutectic phase consistent with metal matrix composite theory \cite{Clyne.93}. Idealized particle morphology minimizes size and aspect ratio, with a consistent distribution. This arrangement maximizes the eutectic particles interfacial surface energy with the surrounding matrix.

\begin{figure}[]
\centering
\includegraphics[width=0.5\linewidth]{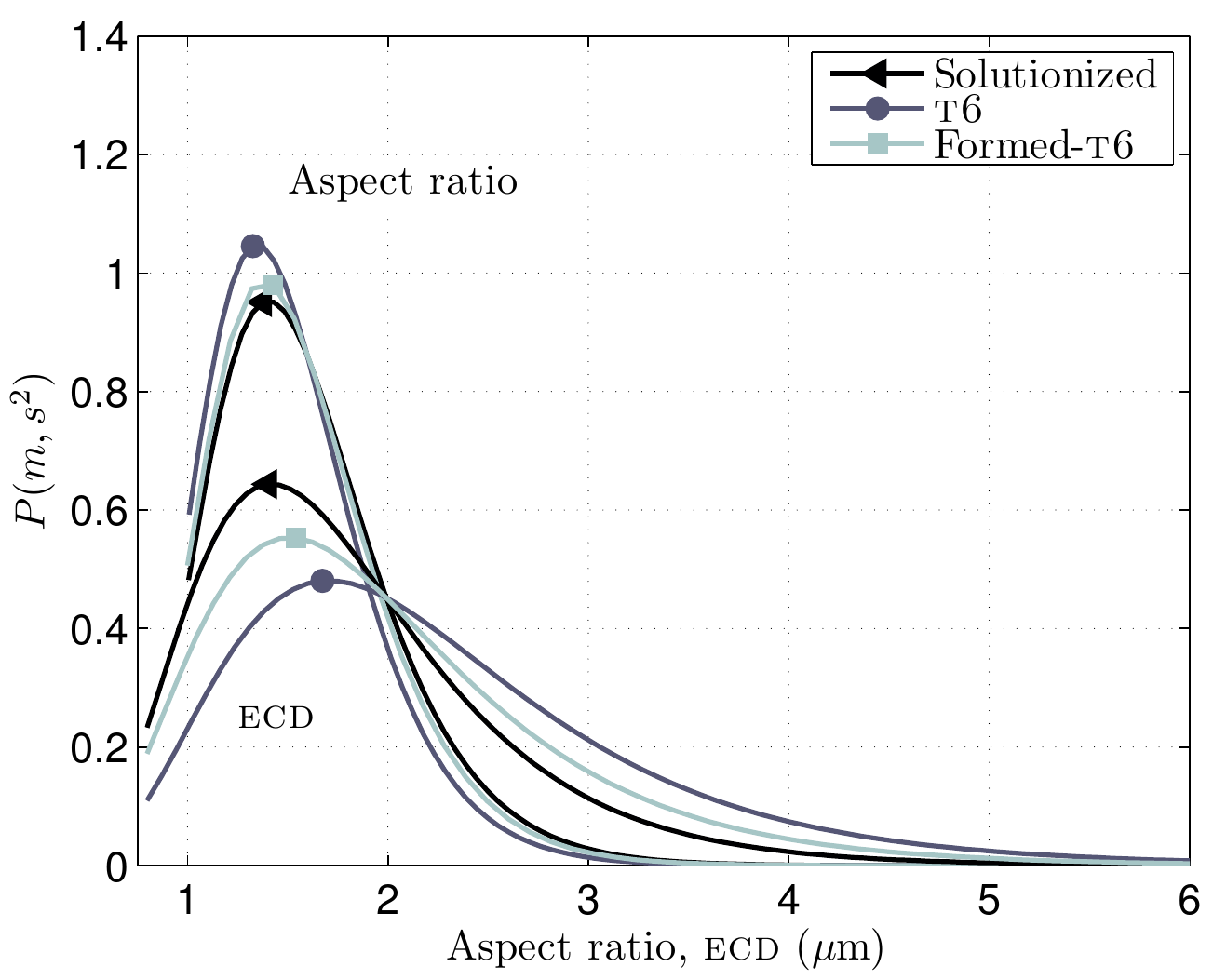}%
\caption[]{Eutectic particle ECD and aspect ratio PDFs.}%
\label{fig:EuStat}
\end{figure}

In order to quantify the effects of different processing paths on the eutectic particles, particle analysis using optical microscopy was conducted on specimens of AC material solutionized at 540$^\circ$C for 50 minutes, unformed-T6 material and T6 treated material drawn from the H specimen. The particle characteristics were quantified with equivalent circle diameter (ECD) and aspect ratios measured from best-fit ellipses. The measurements were then fit to a log-normal probability density function (PDF) according to:
\begin{equation}
P=\frac{1}{x\sqrt{2\pi s^2}}\, \exp\left({-\frac{\left(\ln x-m\right)^2}{2s^2}}\right)
\end{equation}
where $x$ is ECD or aspect ratio, $m$ and $s$ are the mean and standard deviation of the natural logarithm of $x$. The resulting statistics in terms of arithmetic mean, mode, $m$ and $s^2$ are summarized in Table \ref{t:EuStat} and Fig. \ref{fig:EuStat}. This analysis indicates that the aspect ratio does not vary significantly between the different processing paths.  As the melt was chemically identical for all specimens, a possible explanation for this is that the modification technique produces a narrow range of aspect ratios after fragmentation. Wang \cite{Wang.03a} also showed that the distribution of aspect ratio in modified A356/357 was nearly identical, with only unmodified material showing a distinct difference. However, in the present work, a clear difference was noted in ECD, with the smallest values found with solutionized material, then formed-T6 and finally the unformed material having the largest particle size. The ECD and aspect ratio measurements of the unformed material are comparable to those of Wang et al. \cite{Wang.03a} for modified A356-T6, and the ECD measurements are approximately half of those found for unmodified A357-T6 \cite{Wang.03a,Tiryakioglu.03}.

\begin{table}
\caption{Eutectic particle statistics.\label{t:EuStat}}
\centering
    \begin{tabular}{llll}
    \toprule
    \multirow{2}{*}{Material} &  \multirow{2}{*}{Statistic} & \multirow{2}{*}{Aspect ratio} & ECD \\
                             &            &                 & ($\upmu$m)  \\
    \midrule
                                    & Mean  & 1.60     & 1.81     \\
    Solutionized                            & Mode  & 1.40      & 1.42      \\
    (Fig. 11d)        & $m$ &   0.421    &  0.511     \\                                    
                                    & $s^2$ & $8.20\times 10^{-2}$      & 0.162      \\
    \midrule
                                    & Mean  & 1.51     & 2.28     \\
    T6                     & Mode  & 1.35      & 1.70      \\
    (Fig. 11e)        & $m$ &   0.372    &  0.728     \\                                    
                                    & $s^2$ & $7.43\times 10^{-2}$      & 0.195      \\
    \midrule
                                    & Mean  & 1.57     & 2.01     \\
     Formed-T6 & Mode  & 1.38      & 1.51      \\
      (Fig. 12b)      & $m$ &   0.403    &  0.600     \\                      
                                    & $s^2$ & $7.89\times 10^{-2}$      & 0.189      \\
    \bottomrule
\end{tabular}
\end{table}

The results for the solutionized material and the unformed material in the T6 condition are consistent with phenomena associated with standard solution treatment. As reflected in the aspect ratio and ECD measurements, the solutionized material did not coarsen to the same extent as the T6 specimen owing to the longer time at temperature for the latter material. Assuming the $k$ coefficient (Eq. \ref{eq:LSW}) is the same for both formed and unformed material and taking $\bar d$ as the mode value, these results indicate that deformation fragments the eutectic-Si to a much greater extent, leading to smaller eutectic particle sizes after heat treatment. However, it appears that the deformation does not uniformly fragment the eutectic, as the formed materials show marginally larger aspect ratios than unformed.

\section{Phase-specific effects of processing}\label{sec:Micro}
In an attempt to ascertain the degree to which processing history affects the primary versus the eutectic phases, microhardness tests with a very low load were employed to selectively test each phase of material. This was conducted on unformed material with various temperature-time histories, and H specimen material. The unformed material was held at 300, 350, 400 and 540$^\circ$C for 50 minutes, while the formed material was processed over a 25.4 minute period. The results of these measurements, presented in Fig. \ref{fig:microIndent}, show the relative contribution of each phase to the macro hardness and overall strength. The mean microhardness and standard deviation of 30 individual measurements is given for the breadth of conditions presented in the previous section. Indentation locations were chosen such that the plastically affected zone was retained within each phase, as shown in Figs. \ref{fig:microIndent}b and \ref{fig:microIndent}c.


In the AC condition, the eutectic shows a significantly higher hardness as compared to the primary $\upalpha$-Al phase. The decrease in hardness of the primary $\upalpha$-Al phase is identical from AC to hold temperatures of 300 through to 350$^\circ$C, and decreases further to a minimum at 400$^\circ$C. Below holding temperatures of 540$^\circ$C, it appears that the hardness of the eutectic stabilizes after an initial drop, while the peak hardness observed in the primary $\upalpha$-Al phase decreases consistently with temperature. After holding at 540$^\circ$C, both the eutectic and primary $\upalpha$-Al phases increase in microhardness relative to results from holding at lower temperatures. The hardness of the formed material's eutectic is approximately the same as the specimens held below 540$^\circ$C, and the primary $\upalpha$-Al phase is somewhere between the specimens held at 350 and 400$^\circ$C. This indicates that prior to heat-treatment of the formed material, modification to strength for each phase can be attributed to changes in microstructure due to thermal processing.

Following T6 heat treatment, the hardness of both the eutectic and primary $\upalpha$-Al phases increase appreciably, with more effective precipitation and spheroidization. The hardness of the primary $\upalpha$-Al phase in the formed material is approximately the same as the unformed material, with a mean hardness within a standard deviation of the unformed. The mean hardness of the formed material's eutectic phase is 28\% less than that of the unformed, which indicates that the principal cause of the hardness decrease observed in formed samples in the T6 condition (Fig. \ref{fig:All_T6}) is due to changes localized in the eutectic.

\section{Summary}
The hardness and microstructure of A356 following rotary forming at elevated temperatures is affected by a number of different factors across several length scales. Combined hardness profile measurements and microstructural analysis shows that the DAS has less of an effect on hardness than the distribution and condition of eutectic-Si phase. Heating the AC material prior to deformation initiates diffusion-driven coarsening of precipitates and modifies the eutectic structure. An extrapolation of the data from targeted static thermal experiments suggests that the AC material is stable up to approximately 144$^\circ$C. Prior to heat treatment, rotary formed material exhibits decreased macrohardness in-line with the time spent at elevated temperature, indicating that the decrease in hardness between the AC undeformed state to the as-deformed state is principally a thermal effect.  After heat treatment, there was a small macrohardness increase observed in the regions unaffected by forming in the processed material compared to unprocessed material with the same heat treatment. This coincided with a large decrease in macrohardness in heavily deformed regions over unprocessed material, which can be mainly attributed to changes in the eutectic particle morphology, and potential recrystallization.

Eutectic particle size and shape analysis showed that rotary forming fragments the eutectic structure prior to heat treatment. This results in smaller eutectic particles after heat treatment, which was correlated to lower macrohardness. It is therefore surmised that the formed material in the T6 condition may exhibit decreased yield strength as compared to undeformed material in the same state, despite smaller eutectic particles observed in the deformed material. On the basis of interfacial energy, fewer large eutectic particles are not as effective in strengthening as many small particles due to diminished interfacial surface area. However, this assumes uniform distribution and morphology, which is not the case for this material. Barring any evidence of recrystallization, rotary formed parts may require an extended solutionizing treatment to arrive at equivalent strength after heat treatment. In order to fully optimize the heat treatment process for A356 components processed by rotary forming, further study is required to identify the optimum particle size and solutionizing time to arrive at the desired strength and characterizing any recrystallization phenomena which may have occurred.

\end{linenumbers}

\section*{References}
\singlespacing

\end{document}